\documentclass[a4paper,11pt]{article}

\usepackage{amsmath,amssymb}
\usepackage{cite}
\usepackage{hyperref}

\def \be {\begin{equation}}
\def \ee {\end{equation}}
\def \ba {\begin{array}}
\def \ea {\end{array}}
\def \bea{\begin{eqnarray}}
\def \eea{\end{eqnarray}}

\def \G {\Gamma}

\def \m {\mu}

\def \s {\sigma}

\def \o {\omega}
\def \O {\Omega}

\def \p {\partial}
\def \f {\frac}

\def \nn {\nonumber}

\def \ma {\mathcal}

\def \lt {\left}
\def \rt {\right}

\def \lra {\leftrightarrow}
\def \sr {\sqrt}

\def \hs {\hspace}
\def \pp {\propto}

\title{Thermodynamics in Black-hole/CFT Correspondence}
\author{
Bin Chen$^{1,2,3}$\footnote{bchen01@pku.edu.cn},\,
and
Jia-ju Zhang$^{1,2}$\footnote{jjzhang@pku.edu.cn}\,
}
\date{}

\begin{document}

\maketitle

\begin{center}
{\it
$^{1}$Department of Physics, Peking University, Beijing 100871, P.R. China\\
\vspace{2mm}
$^{2}$State Key Laboratory of Nuclear Physics and Technology, Peking University, Beijing 100871, P.R. China\\
\vspace{2mm}
$^{3}$Center for High Energy Physics, Peking University, Beijing 100871, P.R. China\\
}
\vspace{10mm}
\end{center}

\begin{abstract}\noindent
The area law of Bekenstein-Hawking entropy of the black hole suggests that the black hole should have a lower-dimensional holographic description. It has been found recently that a large class of rotating and charged black holes could be holographically described a two-dimensional (2D) conformal field theory (CFT). We show that the universal information of the dual CFT, including the central charges and the temperatures, is fully encoded in the thermodynamics laws of both outer and inner horizons. These laws, characterizing how the black hole responds under the perturbation, allows us to read different dual pictures
with respect to different kinds of perturbations.  The remarkable effectiveness of this thermodynamics method suggest that the inner horizon could play a key role in the study of holographic description of the black hole. 
\end{abstract}

~

\begin{center}
{\it Essay written for the Gravity Research Foundation 2013 Awards for Essays on Gravitation} \\ ~ \\
\today
\end{center}

\newpage

One of central issues in  quantum gravity is to understand the thermodynamics in black hole physics \cite{Bardeen:1973gs}, in particular the Bekenstein-Hawking entropy of a black hole \cite{Bekenstein:1973ur,Hawking:1974sw}. The unexpected area law of the entropy posed a serious problem on how to count the black hole entropy in a microscopic way. In string theory the entropies of some kinds of extreme black holes could be reproduced microscopically by counting the degeneracy of branes configurations forming the black hole \cite{Strominger:1996sh}. In the past few years, the holographic 2D CFT descriptions of usual rotating and charged black holes, which could not be embedded into string theory in a simple way, has been proposed and tested \cite{K1,K2}. In this so-called Kerr/CFT correspondence, people applied the asymptotic symmetry group (ASG) analysis \cite{Brown:1986nw} for the near horizon geometry of extremal black hole \cite{K1}, and studied the hidden conformal symmetry (HCS) in the low-frequency scalar scattering off the non-extreme black hole \cite{K2} to establish the holographic pictures.

One remarkable feature in the holographic description of Kerr and multi-charged black holes is that the central charges of the dual CFT are independent of the masses of the black holes, which could be related to the fact that the area product of the horizons of these black holes are mass-independent \cite{KN}.
For general 4D and 5D multi-charged rotating black holes, the outer and inner horizon entropies could be written respectively as
\be \label{j15}
S_\pm = 2\pi (\sr{N_L} \pm \sr{N_R}),
\ee
where $N_L,N_R$ could be interpreted as the levels of the left- and right-moving excitations of a two-dimensional CFT \cite{Larsen:1997ge,Cvetic:1997uw,Cvetic:1997xv}.
Then the entropy product
\be
S_+S_-=4\pi^2(N_L-N_R)
\ee
should be quantized  due to the level matching condition in CFT.  As a result, the entropy product $S_+S_-$ must be mass-independent, being expressed solely in terms of quantized angular momenta and charges. It was suggested in \cite{Cvetic:2009jn,Cvetic:2010mn} that the mass-independence of the entropy product could be taken as a criterion if a black hole has a holographic picture.

In all the investigations on the area product of the horizons, the thermodynamics of the inner horizon plays an important role. In studying the black hole physics,  the presence of the inner event horizon has often been ignored. There are good reasons to do so: it is inaccessible to an external observer, and moreover it could be unstable. Nevertheless, the first law  of the Kerr black hole inner mechanics has been discussed in \cite{Curir1979} long time ago. Very recently there are growing interests to re-investigate this issue\cite{Castro:2012av,Detournay:2012ug,Chen:2012mh}, especially its implication in setting up the holographic pictures of black holes\cite{Chen:2012mh,Chen:2012ps,Chen:2012pt,Chen:2012yd,Chen:2013rb,Chen:2013aza}.

For stationary rotating charged black holes, the inner horizon thermodynamics is quite simple \cite{Chen:2012mh}. Without losing generality, let us consider 4D Kerr-Newman black hole with two conserved charges, one angular momentum and one electric charge. There is a symmetry between the physical quantities under the exchange of two horizons:
\bea
&&T_-=-T_+|_{r_+\lra r_-}, ~~~
S_-=S_+|_{r_+\lra r_-}, \nn\\
&&\O_-=\O_+|_{r_+\lra r_-}, ~~~
\Phi_-=\Phi_+|_{r_+\lra r_-},  \label{exchange}
 \eea
where $T_\pm$, $\O_\pm$ and $\Phi_\pm$ are the Hawking temperatures, angular velocities and electric potentials of the outer and inner horizons respectively.
This property suggests that once the first law of thermodynamics holds at the outer horizon, so does it at the inner horizon.  In this case, the thermodynamics laws at the outer and inner horizon are respectively 
\bea
&& d M=T_+ d S_+ + \O_+^1 d N_1 + \O_+^2 d N_2,\nn\\
&& d M=-T_- d S_- + \O_-^1 d N_1 + \O_-^2 d N_2.\label{in}
\eea
Here we define $N_2=Q/e$ to be an integer, with $e$ being the unit charge in the Maxwell theory. Such quantization condition is due to the fact that the infalling particle always carries  integer units of
charges. Also we have  made the definitions $\O_\pm^1=\O_\pm$, $N_1=J$, $\O_\pm^2=e\Phi_\pm$ for convenience.
With the above thermodynamics laws, it is easy to show that the entropy product being mass-independence is equivalent to the relation $T_+S_+=T_-S_-$. When the entropy product is mass-independent, we may define
\be \label{e3}
\ma F \equiv \f{S_+ S_-}{4\pi^2},
\ee
which is a function of the charges.

Furthermore, the relations (\ref{j15}) suggest us to compose the left- and right-moving entropies as
\be
S_{R,L}=\f{1}{2}(S_+ \mp S_-).
\ee
Correspondingly one may define various left- and right-moving quantities\cite{Cvetic:1997uw,Cvetic:1997xv,Cvetic:2009jn}
\bea
&& T_{R,L}=\f{T_-T_+}{T_- \pm T_+},  \nn\\
&& \O_{R}^{i}=\f{T_- \O_+^{i} + T_+ \O_-^{i}}{2(T_- + T_+)},  \nn\\
&& \O_{L}^{i}=\f{T_- \O_+^{i} - T_+ \O_-^{i}}{2(T_- - T_+)},\hs{3ex}i=1,2,
\eea
in terms of which the thermodynamics laws  for the right- and left-moving sectors are of the forms
\bea \label{e8}
\f{1}{2}d M&=&T_R d S_R+\O_R^1 d N_1  +\O_R^2 d N_2 \nn\\
\f{1}{2}d M&=&T_L d S_L  +\O_L^1 d N_1  +\O_L^2 d N_2.
\eea
This fact is reminiscent of a  2D CFT, whose left- and right-moving sectors are independent. Therefore one can take this fact
seriously, and relate the two sectors to the sectors of dual CFT.

The first consequence of the above identification is that  the central charges of two sectors of dual CFT must be equal.
Taking into account of the fact that $T_+S_+=T_-S_-$, we get
\be
\frac{S_L}{S_R}=\frac{T_L}{T_R}. \label{ratio}
\ee
From the Cardy formula
\be \label{e6}
S_{R,L}=\f{\pi^2}{3}c_{R,L} T_{R,L}^m,
\ee
we find that 
\be
c_R=c_L.
\ee
Note that the above derivation relies on the fact that the microscopic temperatures $T_{R,L}^m$ of dual CFT are proportional to the temperatures $T_{R,L}$.
 Recall that the thermodynamics relations (\ref{in}) tell us how the black hole responds to the perturbations. Firstly we consider the neutral perturbation with only angular momentum. In this case, keeping $N_2$ invariant, from the above thermodynamics laws we  get
\be \label{e2}
d N_1=T_L^1 d S_L-T_R^1 d S_R,
\ee
with
\be
T_{R,L}^1=T_{R,L}R_1, ~~~ R_1=\f{1}{\O_R^1-\O_L^1}. \label{micT}
\ee
We would like to identify $T_{R,L}^1$ as the right- and left-moving temperatures in the dual CFT. This identification turns out to be correct as these temperatures are exactly the ones read from the hidden conformal symmetry in the low frequency neutral scalar scattering\cite{Chen:2010xu}. The length $R_1$ could be understood as the size of the circle in which the microscopic CFT resides. With the microscopic temperature (\ref{micT}) and using the Cardy formula,  we obtain
\be \label{c1}
c_{R,L}^1=6\f{\p \ma F}{\p N_1},
\ee
which reproduce the central charges for the $J$-picture CFT\cite{Hartman:2008pb}.

We could actually get more from the thermodynamics\cite{Chen:2012ps}. In the $J$-picture we have the perturbation $dM=\o, dN_1=k_1,dN_2=0$ around the black hole such that the thermodynamics laws give
\bea
&& T_R^1 d S_R= R_1 \lt( \f{1}{2}\o-\O_R k \rt),  \nn\\
&& T_L^1 d S_L= R_1 \lt( \f{1}{2}\o-\O_L k \rt).
\eea
On the CFT side, we expect that
\bea
&& T_R^1 d S_R=\o_R^1-q_R^1 \m_R^1,  \nn\\
&& T_L^1 d S_L=\o_L^1-q_L^1 \m_L^1,
\eea
with $\o_{R,L}^1$, $q_{R,L}^1$ and $\m_{R,L}^1$ as the frequencies, charges, and chemical potentials of the operator dual to the perturbation.
 Therefore we find the identifications
\be \label{e31}
\o_{R,L}^1=\f{R_1}{2}\o, ~~~
q_{R,L}^1=k_1, ~~~
\m_{R,L}^1=R_1 \O_{R,L}^1
\ee
They are exactly the same quantities appearing in the absorption cross section of the low-frequency scattering of a neutral scalar\cite{Chen:2010xu,Chen:2010ni}
\bea
&&\s \pp \sinh \lt(\f{ \o_L^1-q_L^1 \m_L^1}{2T_L^1}+\f{ \o_R^1-q_R^1\m_R^1}{2T_R^1} \rt)  \nn\\
&&\phantom{\s\pp} \times  \lt| \G \lt( h_L^1+i \f{\o_L^1-q_L^1\m_L^1}{2\pi T_L^1} \rt) \rt|^2
                           \lt| \G \lt( h_R^1+i \f{\o_R^1-q_R^1\m_R^1}{2\pi T_R^1} \rt) \rt|^2.
\eea

On the other hand, we may consider the charged perturbation without angular momentum. In this case, we should keep $N_1$ invariant and get
 \be
 dN_2=T_L^2dS_L-T^2_RdS_R,
 \ee
 from which we read  the temperatures and the central charges of $N_2$ picture CFT
\bea
&& T_{R,L}^2=\f{T_- \mp T_+}{\O_-^2-\O_+^2},  \label{t2}\\
&& c_{R,L}^2=6\f{\p \ma F}{\p N_2}.\label{c2}
\eea
Note that the temperatures and the central charges obtained here are slightly different from the ones in \cite{Chen:2010ywa} by a factor. Such a factor is ambiguous in \cite{Chen:2010ywa}, but is completely fixed by the quantization condition in our treatment \cite{Chen:2012ps}.

Besides the above two elementary CFT pictures, there could be other general CFT pictures whose universal information could be read in a simple way. The key point is to consider the response of the black hole with respect to the perturbation carrying both angular momentum and electric charge.
For a general perturbation labelled as $(dN_1,dN_2)$, it may carry $dN_1$ units of angular momentum and $dN_2$ units of electric charge. All the perturbations can be classified by two coprime integers $(a,b)$.  For $(a,b)$-type perturbations $(d N_1,d N_2)=dN(a,b)$,  the thermodynamics laws are
\bea
&&\f{1}{2} d M=T_R d S_R + \O_R^N d N \nn\\
&&\f{1}{2} d M=T_L d S_L + \O_L^N d N,
\eea
with
\be
\O_{R,L}^N=a \O_{R,L}^1 + b \O_{R,L}^2.
 \ee
 Using the similar procedure, we obtain a new holographic 2D CFT picture with the temperatures and the central charges
\bea \label{t3}
&& T_{R,L}^{(a,b)}=\f{1}{a/ T_{R,L}^1+b/T_{R,L}^2},  \nn\\
&& c_{R,L}^{(a,b)}=a c_{R,L}^1+b c_{R,L}^2.
\eea
Remarkably, they exactly agree with the results obtained in other conventional ways\cite{Chen:2011wm,Chen:2011kt}.

Let us  summarize the main points of the thermodynamics method in setting up the holographic picture of the black hole:
\begin{itemize}
   \item For Einstein-Maxwell-type theories, the mass-independence of the entropy product function ${\ma F}$, or equivalently the relation $T_+ S_+=T_-S_-$, could be taken as the criterion for a black hole to have a holographic description. If it holds, the thermodynamics laws suggest that central charges of two sectors   of dual CFT must be equal.
   \item The holographic picture could be read from the response of the black hole with respect to the perturbation. The response is encoded in the thermodynamics laws at both the outer and inner horizons. Different kinds of perturbations may give different dual pictures. For each elementary dual $N_i$ picture, which corresponds to the perturbation carrying purely $N_i$-type charge, the central charge is
       \be
       c^i_{R,L}=6\f{\p \ma F}{\p N_i},
       \ee
       which agree exactly with the one obtained from ASG analysis\cite{Chen:2013rb}. On the other hand, a probe scattering off the black hole, especially at the low frequency limit in the near region, can tell us holographic information of the black hole as well\cite{Castro:2013kea}. In a quite similar way, different probes may read out different dual pictures\cite{Chen:2012ps}. It turns out that two different methods are consistent, though the underlying pictures are complementary.
   \item For general pictures, one has to consider the perturbation carrying not only one kind of charge. The corresponding central charge is the linear combination of the central charges of elementary pictures.
   \item The effectiveness of the thermodynamics method is remarkable. It not only allows us to read the temperatures and central charges of dual pictures, it also helps to determine the frequencies, charges and chemical potentials of the operators dual to the perturbation.

\end{itemize}

Thermodynamics of the black hole has being studied for almost forty years, but it still brings us new surprise. It reflects not only the holographic nature of quantum gravity, but also encodes in itself the information of the holographic pictures and the symmetry among them. The recent study indicates that the inner horizon thermodynamics unexpectedly play an indispensable role in establishing the holographic pictures. It certainly deserves further investigations, especially from microscopical points of view.

\vspace*{5mm}
\noindent {{\bf Acknowledgments}}
The work was in part supported by NSFC Grant No. 11275010. JJZ was also in part supported by Scholarship Award for Excellent Doctoral Student granted by Ministry of Education of China.

\end{document}